%% file: main.tex
\documentclass{aa}
\usepackage[varg]{txfonts}
\usepackage{balance}
\usepackage[autostyle, english=british]{csquotes}
\usepackage{pgf}
\usepackage{hyperref}
\usepackage{cleveref}
\Crefname{appendix}{Appendix}{Appendices}
\crefname{appendix}{Appendix}{Appendices}
\usepackage{siunitx}
\usepackage{booktabs}
\usepackage{float}
\usepackage{amsmath, amssymb, bbold, mathrsfs, nicefrac, dsfont, mathtools}
\renewcommand{\Re}{\mathfrak{Re}}
\renewcommand{\Im}{\mathfrak{Im}}
\newcommand{\iu}{{i\mkern1mu}}
\DeclarePairedDelimiter\floor{\lfloor}{\rfloor}
\DeclarePairedDelimiter\ceil{\lceil}{\rceil}
\DeclareMathOperator{\supp}{supp}
\newcommand{\norm}[1]{\big\lVert#1\big\rVert}

\begin{document}

\title{Efficient wide-field radio interferometry response}
\author{Philipp Arras\inst{\ref{mpa},\ref{lmu},\ref{tum}}\and Martin Reinecke\inst{\ref{mpa}} \and Rüdiger Westermann\inst{\ref{tum}}\and Torsten A. Enßlin\inst{\ref{mpa},\ref{lmu}}}
\institute{Max-Planck Institut für Astrophysik, Karl-Schwarzschild-Str. 1, 85748 Garching, Germany\label{mpa}\and Ludwig-Maximilians-Universität München (LMU), Geschwister-Scholl-Platz 1, 80539 München, Germany\label{lmu}\and Technische Universität München (TUM), Boltzmannstr. 3, 85748 Garching, Germany\label{tum}}
\date{Received <date>/ Accepted <date>}
\abstract{
  Radio interferometers do not measure the sky brightness distribution directly, but measure a modified Fourier transform of it.
  Imaging algorithms therefore need a computational representation of the linear measurement operator and its adjoint, regardless of the specific chosen imaging algorithm.
  In this paper, we present a C++ implementation of the radio interferometric measurement operator for wide-field measurements that is based on so-called improved \textit{w}-stacking.
  It can provide high accuracy (down to $\approx 10^{-12}$),
  is based on a new gridding kernel that allows smaller kernel support for given accuracy,
  dynamically chooses kernel, kernel support, and oversampling factor for maximum performance,
  uses piece-wise polynomial approximation for cheap evaluations of the gridding kernel,
  treats the visibilities in cache-friendly order,
  uses explicit vectorisation if available,
  and comes with a parallelisation scheme that scales well also in the adjoint direction (which is a problem for many previous implementations).
  The implementation has a small memory footprint in the sense that temporary internal data structures are much smaller than the respective input and output data,
  allowing in-memory processing of data sets that needed to be read from disk or distributed across several compute nodes before.
}
\maketitle


\section{Introduction}
The central data analysis task in radio interferometry derives the location-dependent sky brightness distribution $I(l,m)$ from a set of complex-valued measured visibilities $d_k$.
In the noise-less case they are related by the expression \citep[e.g.][]{book}
\begin{align}
\label{eq:measurement}
d_k = \iint \frac{e^{2\pi\iu\lambda_k^{-1} \tilde w_k(n(l,m)-1)}}{n(l,m)}\, I(l,m)\, e^{-2\pi \iu\lambda_k^{-1} (\tilde u_kl+\tilde v_km)}   \,dl\, dm .
\end{align}
Here, $l$, $m$, and $n\coloneqq \sqrt{1-l^2-m^2}$ are direction cosines with respect to the central observation axis, while $\tilde u_k$, $\tilde v_k$ , and $\tilde w_k$ are the coordinates of the baselines in metres and $\lambda_k$ are the observation wavelengths.
When we assume that $I(l,m)$ is approximated by discretised values on a Cartesian ($l$, $m$) grid, the double integral corresponds to a discrete Fourier transform.
The entries $d_k$ of the data vector $d$ correspond to delta peak readouts of the three-dimensional Fourier transformed sky brightness at Fourier location $(u_k, v_k, w_k)$, which are commonly called \enquote{visibilities}.
It suffices to discuss the noise-less case here.
While taking the noise into account is the task of the chosen imaging algorithm, all such algorithms need an implementation of \cref{eq:measurement}.

Typical problem sizes range from $10^6$ to beyond $10^9$ visibilities, fields of view can reach significant fractions of the hemisphere, and image dimensions exceed $10\,000\times 10\,000$~pixels.
It is evident that na{\"\i}ve application of \cref{eq:measurement} becomes prohibitively expensive at these parameters; a single evaluation would already require $\approx 10^{17}$ calls to the complex exponential function.

Massive acceleration can be achieved by using \enquote{convolutional gridding} \citep[in other fields often called \enquote{non-uniform fast Fourier transform};][]{nfft}.
Here, the information contained in the $d_k$ is transferred onto a regular Cartesian grid by convolving the delta peak readouts at $(u_k, v_k, w_k)$ with an appropriately chosen kernel function, which is evaluated at surrounding $(u,v)$ grid points.
Transformation between $u,v$ and $l,m$ can now be carried out quickly by means of a two-dimensional fast Fourier transform \citep[FFT; ][]{fft}, and the smoothing caused by the convolution with the kernel is compensated for by dividing the $I(l,m)$ by the Fourier-transformed kernel.

When the term $e^{-2\pi\iu \lambda^{-1} \tilde w(n-1)}/n$ is very close to 1, no further optimisation steps are required.
This criterion
is not fulfilled for non-planar instruments and for wide-field observations.
Therefore the visibilities need to be gridded onto several $uv$-planes with different $w$, which are Fourier-transformed and corrected separately.
\citet{ricknoncoplanar} has pointed out that \cref{eq:measurement} can be written as a three-dimensional Fourier transform.
Based on this idea, \citet{wgridding} applied the convolutional gridding algorithm not only for the $uv$-coordinates, but also for the $w$-direction.
Because this approach naturally generalises $w$-stacking \citep{wsclean} to use gridding in the $w$-direction as well,
we propose the term \enquote{$w$-gridding} instead of the term \enquote{improved $w$-stacking} \citep{wgridding,ye21}.

This paper does not present any new
insights into the individual components of the radio interferometric measurement operator implementation  (except for the introduction of a tuned gridding kernel in \cref{sec:kernelshape});
our code only makes use of algorithms that are already publicly available.
Instead, our main point is to demonstrate how significant advances in performance and accuracy can be achieved by appropriate selection of individual components and their efficient implementation.
Our implementation has been integrated into the well-known imaging tool \texttt{wsclean}\footnote{\url{https://gitlab.com/aroffringa/wsclean}} \citep{wsclean} since version2.9, where it can be selected through the \texttt{-use-wgridder} flag, and the imaging toolkit \texttt{codex-africanus}\footnote{\url{https://github.com/ska-sa/codex-africanus}}.
Furthermore, the implementation presented here has been used in \citet{eht, rickvsresolve}, for instance.

\Cref{sec:math} introduces the notation used in this paper and summarises the algorithmic approach to numerically approximate \cref{eq:measurement} and its adjoint.
\Cref{sec:algo} describes all algorithmic components in detail from a computational point of view, and \cref{sec:impl} lists the design goals for the new code, which influence the choice of algorithmic components from the set given in \cref{sec:algo}.
Here we also list a number of additional optimisations to improve overall performance.
The new code is validated against discrete Fourier transforms in \cref{sec:validation}, and an analysis of its scaling behaviour as well as a performance comparison with other publicly available packages is presented in \cref{sec:benchmarks}.

\section{Notation and formal derivation of the algorithm}\label{sec:math}
The data that are taken by radio interferometers are called \enquote{visibilities}.
\Cref{eq:measurement} already shows that the operation that is to be implemented is similar to a Fourier transform modulated by a phase term.
In the following, we introduce all notation that is required to describe the algorithm and present the three-dimensional (de)gridding approach from \citet{wgridding} in this notation.

Let $\lambda \in \mathbb R^{n_\nu}$ be the vector of observing wavelengths in metres and $(\tilde u, \tilde v, \tilde w)$ the coordinates of the baselines in metres, each of which are elements of $\mathbb R^{n_r}$.
In other words, $n_\nu$ and $n_r$ are the number of observing wavelengths and number of rows of the data set, respectively.
Then, the effective baseline coordinates $(u, v, w)$ are defined as
\begin{align}
u \coloneqq \tilde u\otimes \lambda^{-1},\quad v \coloneqq \tilde v\otimes \lambda^{-1},\quad w \coloneqq \tilde w\otimes \lambda^{-1}.
\end{align}
These are the effective locations of the sampling points in Fourier space.
To simplify the notation, we view the above three coordinates as elements of a simple vector space, for example,\ $u\in\mathbb R^{n_d}$ with $n_d = n_rn_\nu$.
Because the measurement equation (\ref{eq:measurement}) is to be evaluated on a computer, it needs to be discretised,
\begin{align}
  \label{eq:rime}
  (R_0I)_k \coloneqq \sum_{l\in L}\sum_{m\in M} \frac{e^{-2\pi \iu [u_k l + v_k m - w_k(n_{lm}-1)]} \, I_{lm}}{n_{lm}},\quad k\in\{0, \ldots, n_d-1\},
\end{align}
where $R_0$ is the (accurate) response operator defined by the right-hand side of the equation, and $L, M$ are the sets of direction cosines of the pixels of the discretised sky brightness distribution in the two orthogonal directions.
$(\Delta l, \Delta m)$ are the pixel sizes, and  $(n_l, n_m)$ are the number of pixels.
Then, formally, $L$ and $M$ can be defined as
\begin{align}
  L &\coloneqq \Big\{ (-\tfrac{n_l}{2} + j)\,\Delta l \:|\: j \in \{0, \ldots, n_l-1\}\Big\},\\
  M &\coloneqq \Big\{ (-\tfrac{n_m}{2} + j)\,\Delta m \:|\: j \in \{0, \ldots, n_m-1\}\Big\}.
\end{align}
It is apparent that computing \cref{eq:rime} is prohibitively expensive because the whole sum needs to be performed for each data index $k$ individually.
As a solution, the convolution theorem can be applied in order to replace the Fourier transform by an FFT that can be reused for all data points.
As it stands, \cref{eq:rime} is not a pure Fourier transform because of the phase term $w_k (n_{lm}-1)$.
As discussed above, we follow \citet{ricknoncoplanar} and introduce an auxiliary Fourier integration in which $w$ and $n_{lm}-1$ are viewed as Fourier-conjugate variables,
\begin{align}
  \label{eq:rime3d}
 (RI)_k =\sum_{l\in L}\sum_{m\in M} \int_{\tilde n \in \mathbb R} e^{-2\pi\iu [u_k l + v_k m + w_k \tilde n]}\, \frac{ \delta ( \tilde n - (1 - n_{lm} ))}{n_{lm}}\, I_{lm}\, d \tilde n .
\end{align}
The next goal is to replace the above three-dimensional non-equidistant Fourier transform by an equidistant one.
This can be done by expressing a visibility $d_k$ as a convolution of a field defined on a grid with a convolution kernel.
This convolution is undone by dividing by its Fourier transform in sky space.

For this, we need to define the convolution kernel.
Let $\phi : \mathbb R \to \mathbb R^+$ be a function that is point-symmetric around 0 and has compact support $\supp (\phi) = [-\tfrac\alpha2, \tfrac\alpha2] $ with the kernel support size \mbox{$\alpha \in \mathbb N$}.
In other words, the kernel function is zero outside a symmetric integer-length interval around zero.
In practice, this means that every visibility is gridded onto the $\alpha\times \alpha$ $uv$-grid points that are closest to it.
We use $\phi$ as convolution kernel to interpolate all three grid dimensions.
Let $\psi : [-\tfrac12, \tfrac12] \to \mathbb R$ be its Fourier transform: $\psi(k) \coloneqq \int_{-\infty}^\infty \phi(x)e^{\iu k x }\, dx$.
$\psi$ needs to be defined only on $[-\tfrac12, \tfrac12]$ because $\phi$ is evaluated on a grid with pixel size 1.

Now, the (discrete) convolution theorem can be applied to turn the sums in \cref{eq:rime3d} into a discrete Fourier transform followed by a periodic convolution on an oversampled grid (with oversampling factor $\sigma$) and to turn the integral over $\tilde n$ into a regular convolution.
Some degree of oversampling ($\sigma>1$) is required to lower the error of the algorithm to the required levels; ultimately, the error depends on $\sigma$, $\alpha$, and the kernel $\phi$.
Specifically for the $w$-direction, using the coordinate transform $c(x) = w_k+x\Delta w$ and the definition of $\psi$,
\begin{align}
  e^{2\pi \iu (n_{lm}-1) w_k} \psi\Big([n_{lm}-1]\Delta w\Big) &= \int_{-\infty}^\infty e^{2\pi\iu (n_{lm}-1) (w_k + \Delta w x)} \phi(x) \, dx \\
  &= \int_{-\infty}^\infty e^{2\pi \iu (n_{lm}-1) c} \phi \Big(\tfrac{c-w_k}{\Delta w}\Big)\, \frac{dc}{\Delta w}\\
  &\approx \sum_{c\in W} e^{2\pi \iu (n_{lm}-1) c} \phi \Big(\tfrac{c-w_k}{\Delta w}\Big),\label{eq:approxw}
\end{align}
with $W = \Big\{w_\circ + j\Delta w \:\big|\: j\in\{0, \ldots, N_w -1\}\Big\}$, it follows that
\begin{align}
  e^{2\pi \iu (n_{lm}-1) w_k} \approx \frac{\sum_{c\in W} e^{2\pi \iu (n_{lm}-1) c} \phi \Big(\tfrac{c-w_k}{\Delta w}\Big)}{\psi\Big([n_{lm}-1]\Delta w\Big)}.\label{eq:approxw2}
\end{align}
This expression replaces the $w$-term in \cref{eq:rime3d} below.
The idea of rewriting the $w$-term as a convolution was first presented in \citet{wgridding}.
$w_\circ$, $N_w$ , and $\Delta w$ denote the as yet unspecified position of the $w$-plane with the lowest $w$-value, the number of $w$-planes, and the distance between neighbouring $w$-planes, respectively.
The approximation \cref{eq:approxw} is only sensible for all $l \in L, m\in M$ and all $k$ if $\Delta w$ is small enough.
The proper condition is given by the Nyquist-Shannon sampling theorem \citep{wgridding},
\begin{align}
\max_{(l, m) \in L\times M}  2\Delta w \,\sigma |n_{lm}-1| \leq 1.
\end{align}
The factor $\sigma$ appears because the accuracy of a given gridding kernel $\phi$ depends on the oversampling factor.
Therefore, the optimal, that is,\ largest possible, $\Delta w$ is
\begin{align}
  \Delta w = \min_{(l, m) \in L\times M} \frac{1}{ 2 \sigma |n_{lm}-1|}.
\end{align}
For a given $\sigma$ this determines $\Delta w$.
$w_\circ$ and $N_w$ are still unspecified.
Combining \cref{eq:rime3d} and \cref{eq:approxw2} leads to the final approximation of the measurement equation,
\begin{align}
  \label{eq:nfft}
  (R I)_k \coloneqq \sum_{a\in U}\sum_{b\in V}\sum_{c\in W} \Phi_k(a,b,c) \sum_{l\in L}\sum_{m\in M} e^{-2\pi\iu [a l + b m + c(n_{lm} - 1)]}\,\frac{ I_{lm}}{n_{lm}\Psi_{lm}},
\end{align}
where $R$ is the linear map that approximates $R_0$ in our implementation, and $\Phi$ and $\Psi$ are the threefold outer product of $\phi$ and $\psi$, respectively,
\begin{align}
  \Phi_k (a, b, c)  &=  \phi(N_u\,\chi(a-u_k\Delta l))\, \phi(N_v\,\chi(b-v_k\Delta m))\, \phi  \left(\tfrac{c-w_k}{\Delta w}\right),\\
  \Psi_{lm} &= \psi\Big(\tfrac{l}{\sigma  n_x\Delta x}\Big)\, \psi\Big( \tfrac{m}{\sigma n_y\Delta y} \Big)\, \psi\Big([n_{lm}-1]\Delta w\Big),
\end{align}
with $\chi(a) = a - \floor{a} - 0.5$ where $\floor{a} \coloneqq \max \{n\in\mathbb Z \,|\, n \leq a\}$.
To define the sets $U, V$, and $W$, the discretisation in $uvw$-space needs to be worked out.
The number of pixels in discretised $uv$-space is controlled by the oversampling factor $\sigma$,
\begin{align}
  N_l &= \ceil{\sigma n_l},& N_m &= \ceil{\sigma n_m}, &\text{for }\sigma & > 1,
\end{align}
where  $\ceil{a} \coloneqq \min \{n\in\mathbb Z \,|\, n \geq a\}$.
Thus, the set of pixels of the discretised $uvw$-space is given by
\begin{align}
  U &= \Big\{ -\frac12 + \frac{j}{N_u} \:\big|\: j \in \{ 0, \ldots, N_u-1 \} \Big\},\\
  V &= \Big\{ -\frac12 + \frac{j}{N_v} \:\big|\: j \in \{ 0, \ldots, N_v-1 \} \Big\}.
\end{align}

For the $w$-dimension we can assume $w_k \geq 0 $ for all $k$  without loss of generality because the transformation
\begin{equation}
(u_k, v_k, w_k, d_k)\to (-u_k,-v_k,-w_k, d_k^*)\label{eq:Hermitian-symmetry}
\end{equation}
leaves \cref{eq:rime} invariant individually for each $k$.
Because of this Hermitian symmetry, only half of the three-dimensional Fourier space needs to be represented in computer memory.

For a given $\Delta w$, the first $w$-plane is located at
\begin{align}
  \label{eq:wcirc}
 w_\circ = \min_k w_k - \frac{\Delta w \,(\alpha -1)}{2},
\end{align}
that is,\ half of the kernel width subtracted from the minimum $w$-value, and the total number of $w$-planes $N_w$ is
\begin{align}
 N_w = \frac{\max_k w_k - \min_k w_k }{\Delta w} + \alpha\text{,}
\end{align}
because below the minimum and above the maximum $w$-value, half a kernel width needs to be added in order to be able to grid the respective visibilities with extreme $w$-values.

In \cref{eq:nfft}, we can view the sky brightness distribution $I$ as element of $\mathbb R^{n_ln_m}$ and $d \in \mathbb C^{n_k}$.
Then \cref{eq:nfft} can be written as $d = R(I)$ with $R: \mathbb R^{n_ln_m}\to \mathbb C^{n_k}$ being a $\mathbb R$-linear map.
In imaging algorithms this linear map often appears in the context of functionals that are optimised, for example,\ a negative log-likelihood or a simple $\chi^2=|d-R(I)|^2$ functional between data and expected sky response.
To compute the gradient (and potentially higher derivatives) of such functionals, not only $R$, but also $R^\dagger$, the adjoint, is needed.
It can be obtained from \cref{eq:nfft} by reversing the order of all operations and taking the conjugate of the involved complex numbers.
In the case at hand, it is given by
\begin{align}
  \label{eq:adjnfft}
  (R^\dagger d)_{lm} &= \frac{1}{n_{lm}\Psi_{lm}} \sum_{a\in U} \sum_{b\in V} \sum_{c\in W} e^{2\pi\iu [a l + b m + c(n_{lm} - 1)]}\,\sum_k \Phi_k (a, b, c)\, d_k.
\end{align}
Here we can already observe that parallelisation over the data index $k$ is more difficult in \cref{eq:adjnfft} than in \cref{eq:nfft}.
In \cref{eq:adjnfft}, the grid in Fourier space is subject to concurrent write accesses, whereas in \cref{eq:nfft}, it is only read concurrently, which is less problematic.
In \cref{sec:parallel} we discuss this in more detail and present a parallelisation strategy that scales well in both directions.

All in all, the scheme \cref{eq:nfft}, which approximates the discretised version (eq.\ \ref{eq:rime}) of the radio interferometric response function (eq. \ref{eq:measurement}), has been derived.
That it can be computed efficiently is shown in the subsequent sections.
The choice of the gridding kernel function $\phi$, the kernel support $\alpha$, and the oversampling factor $\sigma$ have not yet been discussed.
Tuning these three quantities with respect to each other controls the achievable accuracy and the performance of the algorithm (see \cref{sec:kernelshape}).

\section{Algorithmic elements}\label{sec:algo}

\Cref{eq:nfft} prescribes a non-equidistant Fourier transform that is carried out with the help of the as yet unspecified gridding kernel $\Phi$.
Its choice is characterised by a trade-off between accuracy (larger kernel support $\alpha$ and/or oversampling factor $\sigma$) and computational cost.
As a criterion for assessing the accuracy of a kernel, we use a modified version of the least-misfit function approach from \citet{optimalgridding}.

\subsection{Gridding and degridding and treatment of the $w$-term}
For the implementation, \cref{eq:nfft} is reordered in the following way:
\begin{align}
  (RI)_k &= \sum_{c\in W}\left[  \sum_{a\in U}\sum_{b\in V}\Phi_k(a,b,c) \sum_{l\in L}\sum_{m\in M} e^{-2\pi\iu [a l + b m]}\, \tilde I_{lmc}  \right] \label{eq:R3}\\
  \tilde I_{lmc} &\coloneqq e^{2\pi\iu c (1-n_{lm})}\, \tilde I_{lm}\label{eq:R2}\\
  \tilde I_{lm} &\coloneqq \frac{I_{lm}}{n_{lm}\Psi_{lm}}.\label{eq:R1}
\end{align}
In other words, first the geometric term $n$ and the gridding correction $\Psi$ are applied to the input $I_{lm}$ (eq.\ \ref{eq:R1}).
Then, the $w$-planes are handled one after another.
For every $w$-plane the phase term $e^{2\pi\iu c(1-n_{lm})}$, called $w$-screen, is applied to the image (eq.\ \ref{eq:R2}).
This is followed by the Fourier transform and the degridding procedure with $\Phi$ (bracketed term in \cref{eq:R3}).
Finally, the contributions from all $w$-planes are accumulated by the sum over $c\in W$ to obtain the visibility $d_k$.

For the adjoint direction, \cref{eq:adjnfft} is reordered to
\begin{align}
  (R^\dagger d)_{lm} &= \frac{1}{n_{lm}\Psi_{lm}}  \sum_{c\in W} e^{-2\pi\iu c(1-n_{lm})}\, H_c,\label{eq:Rad1}\\
  H_c &\coloneqq\sum_{a\in U} \sum_{b\in V}  e^{2\pi\iu [a l + b m]}  \sum_k \Phi_k (a, b, c)\, d_k.\label{eq:Rad2}
\end{align}
In words, the $w$-planes are handled one after another again.
First, the visibilities that belong to the current $w$-plane are gridded onto a two-dimensional grid with $\Phi$ and the two-dimensional Fourier transform is applied (eq.\ \ref{eq:Rad2}).
Then, its result $H_c$ is multiplied with the complex conjugate $w$-screen and the contributions from $w$-planes to the image are accumulated by the sum over $c\in W$ (eq.\ \ref{eq:Rad1}).
Finally, the gridding correction $\Psi_{lm}$ and the geometric factor $n_{lm}$ are applied.

The number of iterations in the loop over the $w$-planes $W$ can be reduced by up to a factor of two by restricting the $w$ coordinate to $w \geq 0$ with the help of the Hermitian symmetry (eq.\ \ref{eq:Hermitian-symmetry}).
The implementation scheme described above highlights that the choice of the kernel shape $\phi$ and its evaluation are crucial to the performance of the algorithm:
The support $\alpha$ should be small in order to reduce memory accesses and kernel evaluations.
At the same time, the oversampling factor $\sigma$ needs to be small such that the Fourier transforms do not dominate the run time.
Additionally, the kernel itself needs to be evaluated with high accuracy, while at the same time, its computation should be very fast.

\subsection{Kernel shape}\label{sec:kernelshape}

As already mentioned, the shape of the employed kernel function $\phi$ has a strong effect on the accuracy of the gridding and degridding algorithms.
The historical evolution of preferred kernels is too rich to be discussed here in full, but see \cite{optimalgridding} for an astronomy-centred background and \cite{logsemicircle} for a more engineering-centred point of view.

It appears that the kernel shape accepted as \enquote{optimal} amongst radio astronomers is the spheroidal function as described by \cite{spheroidal}.
This function maximises the energy in the main lobe of the Fourier-transformed kernel compared to the total energy, which is essential to suppress aliasing artefacts.

However, this concept of optimality only holds under the assumption that gridding and degridding are carried out without any oversampling of the $uv$-grid and the corresponding trimming of the resulting dirty image.
While this may have been the default scenario at the time this memorandum was written, most currently employed gridding algorithms use some degree of oversampling and trimming (i.e.\ $\sigma>1$), which requires restating the optimality criterion:
instead of trying to minimise the errors over the entire dirty image, the task now is to minimise the error only in the part of the dirty image that is retained after trimming, while errors in the trimmed part may be arbitrarily high.
More quantitatively: Given a kernel support of $\alpha$ cells and an oversampling factor of $\sigma$, a kernel shape is sought that produces the lowest maximum error within the parts of the dirty image that are not trimmed.

\citet{optimalgridding} demonstrated an approach to determine non-analytic optimal kernels.
However, very good results can also be obtained with rather simple analytical expressions.
\citet{logsemicircle} presented the one-parameter kernel called \enquote{exponential of a semicircle kernel} or \enquote{ES kernel},
\begin{align}
  \phi_\beta: \left[-\tfrac{\alpha}{2},\tfrac{\alpha}{2}\right]  \to  \mathbb R^+,\;
  x \mapsto \exp{ \left(\alpha\beta\left[\sqrt{1-\left(2x/\alpha\right)^2}-1\right]\right)},
\label{eq:semicircle}
\end{align}
for $\beta > 0$.
In the following, we use a two-parameter version derived from this,
\begin{align}
  \phi_{\beta \mu}: \left[-\tfrac{\alpha}{2},\tfrac{\alpha}{2}\right]  \to  \mathbb R^+,\;
  x \mapsto \exp{ \left(\alpha\beta\left[\left( 1-\left(2x/\alpha\right)^2\right)^\mu-1\right]\right)},
\label{eq:semicircle2}
\end{align}
for $\beta > 0$ and $\mu > 0$ and call it \enquote{modified ES kernel}.

To determine optimal values for the two parameters for given $\alpha$ and $\sigma$, we use the prescription described in \citet{optimalgridding}.
The idea is to consider the squared difference between the outputs of the accurate and the approximate adjoint response operator $R_0$ and $R$.
Without loss of generality, we restrict the following analysis to the case of a one-dimensional non-equidistant Fourier transform.
For readability, we define $\tilde \psi(x) \coloneqq \psi\left( \tfrac{x}{\sigma n_x \Delta x} \right)$ and $\tilde \phi_k (a)\coloneqq \phi\left( N_u\, \chi(a-u_k \Delta l) \right)$ and
\begin{align}
  (\tilde R_0^\dagger d)(x) &\coloneqq \sum_k d_k e^{2\pi \iu u_k x},\\
 (\tilde R^\dagger d)(x) &\coloneqq  \tilde \psi(x)^{-1} \, \sum_k d_k \sum_{a\in U} \tilde \phi_k(a)\, e^{2\pi\iu a x}.
\end{align}
Using the Cauchy-Schwarz inequality, the squared error can be bounded from above with
\begin{align}
\left|(\tilde R_0 d-\tilde R d)(x) \right|^2 &=\left| \sum_k d_k e^{2\pi\iu u_k x} \left( 1- \sum_{a\in U} \frac{   e^{2\pi\iu (a-u_k) x}\, \tilde \phi_k(a)}{\tilde \psi(x)}  \right)  \right|^2\\
  & \leq \left( \sum_k \left| d_k \right|^2  \right) \sum_k \left|  1-\sum_{a\in U} \frac{   e^{2\pi\iu (a-u_k) x}\, \tilde \phi_k(a)}{\tilde \psi(x)} \right|^2.
    \label{eq:cauchyschwarz}
\end{align}
The first term of the right-hand side of the inequality is purely data dependent and therefore not relevant in quantifying the (upper limit of the) approximation error of the linear map $R^\dagger$.
The actual approximation error does depend on the data $d$, and for a given data vector, more accurate approximation schemes could be derived in principle.
However, because generic statements about $d$ are difficult to make and a data-independent generic kernel is desired here, we optimise the right-hand side of the inequality.
If the number of visibilities is large (tests have shown that in generic setups already $>10$ visibilities suffice), the values of $(a-u_{k})x \text{ mod } 2\pi$ sample the interval $[0,1)$ sufficiently uniformly.
Then the second term is approximately proportional to the data-independent integral
\begin{align}
  \label{eq:maperror}
  l^2(x) \coloneqq \int_0^1 \left| 1-  \sum_{a\in U}\frac{ e^{2\pi\iu (a-\nu) x}\, \tilde \phi_k(a) }{\tilde \psi(x)}  \right|^2 \, d\nu .
\end{align}
Because the actual error is quantified by $l(x)$, we call $l(x)$ the \enquote{map error function} in contrast to \citet{optimalgridding}, who used this name for $l^2(x)$.
$l(x)$ depends on the choice of the functional form of $\phi$, the kernel support $\alpha$, and the oversampling factor $\sigma$.
\citet{optimalgridding} used \cref{eq:maperror} in a least-squares fashion to determine the \enquote{optimal gridding kernel} for a given $\alpha$ and $\sigma$.

We propose to use \cref{eq:maperror} slightly differently.
Instead of the L2-norm, we use the supremum norm to minimise it because the error should be guaranteed to be below the accuracy specified by the user for all $x$.
Additionally, we use the two-parameter modified ES kernel.
The parameters that result from a two-dimensional parameter search are hard-coded into the implementation.
For explicitness, a selection of parameters is displayed in \cref{app:kernelparams}.

\begin{figure}
  \centering
  \input{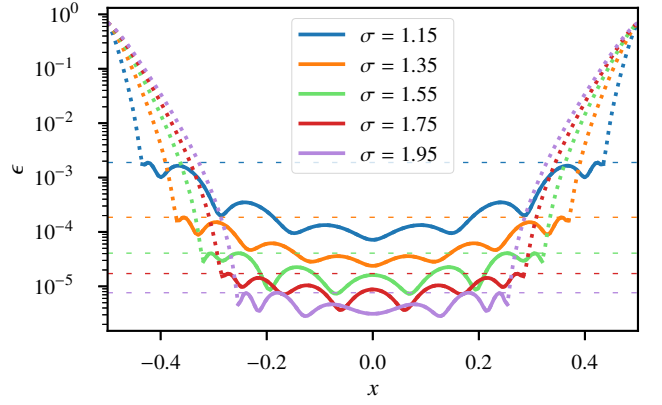}
  \caption{Map error function for kernel support $\alpha = \protect\input{maperrorsupport}$ for a varying oversampling factor $\sigma$.
  The horizontal dotted lines display the advertised accuracy of the kernel.}
  \label{fig:maperror}
\end{figure}

As an example, \cref{fig:maperror} shows the map error function of the modified ES kernel in dependence on the oversampling factor $\sigma$ and for fixed $\alpha$.
Increasing the oversampling factor allows a reduction of the convolution kernel support size while keeping the overall accuracy constant, which reduces the time required for the actual gridding or degridding step.
At the same time, however, an increase in $\sigma$ implies both a larger $uv$-grid and a higher number of $w$-planes.
The former aspect leads to increased memory consumption of the algorithm, and both aspects increase the total cost of FFT operations.
As a consequence, for a given number of visibilities, dirty image size, $w$ range,  and desired accuracy,
it is possible to minimise the algorithm run-time by finding the optimal trade-off between oversampling factor and kernel support size.
The sweet spot for most applications lies in the range 1.2 to 2.0 for the oversampling factor.
Our chosen functional form of the gridding kernel naturally leads to higher accuracy towards the phase centre, that is,\ $x=0$.

\begin{figure}
  \centering
  \input{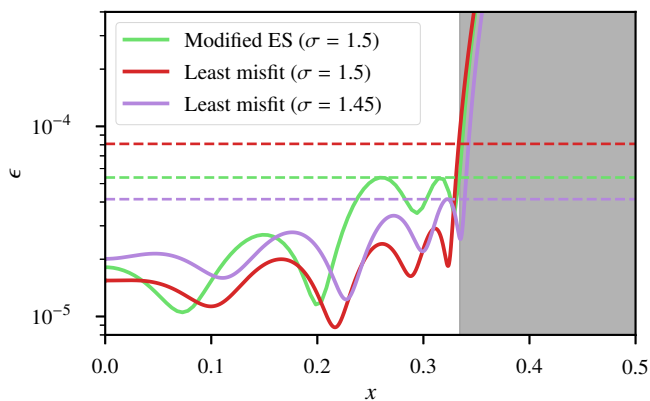}
  \caption{Comparison of the map error function for least-misfit kernels with different oversampling factor and modified ES kernel.
    The kernel support size is $\alpha = \protect\input{mesvskb_alpha}$ for all three kernels.
    The dashed lines denote the supremum norm of the respective functions.
    We display only positive $x$ (in contrast to \cref{fig:comparekernels}).
  All map error functions are symmetric around $x=0$.}
  \label{fig:comparekernelsadd}
\end{figure}

For the comparison of our modified ES kernel and the least-misfit kernel, we note that the kernels are designed to minimise the supremum norm and the L2-norm, respectively, of the map error function.
All least-misfit kernels in the following were computed using the software released alongside \citet{optimalgridding}.
For given $\alpha$ and $\sigma$, the least-misfit kernel is therefore not necessarily optimal in our metric and vice versa, and comparison becomes non-trivial.
\Cref{fig:comparekernelsadd} displays the map error function for the modified ES kernel and the least-misfit kernel with the same $\alpha$ and $\sigma$ and compares it to the least-misfit kernel with $\sigma = 1.45$.
The steep increase in the map error function of the least-misfit kernel for $\sigma = 1.5$ significantly affects the supremum norm but still leads to a lower value for the L2-norm because the function is considerably smaller for small $x$.
For the following comparison we select the least-misfit kernel for $\sigma = 1.45$ by hand.
It is optimal under the L2-norm for $\sigma = 1.45$, but still performs better than the modified ES kernel even at $\sigma =1.5$ under the supremum norm.
It is to be assumed that with a more systematic search, even better least-misfit kernels can be found, so that the selected one should be regarded only in a qualitative sense.

\begin{figure}
  \centering
  \input{img/kernel.pgf}
  \caption{Optimal kernel shapes for $\sigma = \protect\input{mesvskb_sigma}$ and $\alpha = \protect\input{mesvskb_alpha}$ with achieved accuracy $\epsilon$.}
  \label{fig:kernel}
\end{figure}

\begin{figure}
  \centering
  \input{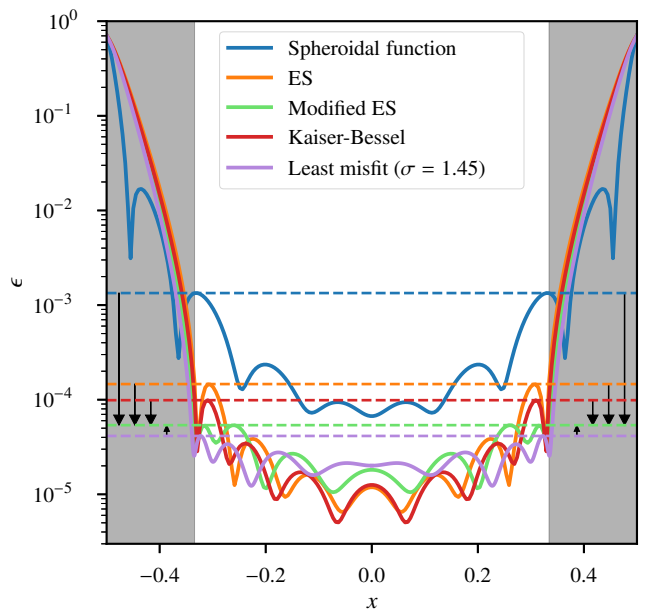}
  \caption{Map error function of different kernel shapes for $\sigma = \protect\input{mesvskb_sigma}$ and $\alpha = \protect\input{mesvskb_alpha}$.
  A least-misfit kernel for a slightly lower oversampling factor is added for qualitative comparison (see the main text for a discussion of this choice), as well as the classic spheroidal function kernel.
  The arrows highlight the differences of the supremum norm of map error function of the different kernels with respect to our modified ES kernel.
  }
  \label{fig:comparekernels}
\end{figure}

\Cref{fig:kernel,fig:comparekernels} display a comparison for given oversampling factor and kernel width of different gridding kernels.
For all kernels (except for the least-misfit kernel) the same hyperparameter search for optimal parameters given $\alpha$ and $\sigma$ was performed.
The ES kernel \citep{logsemicircle} is less accurate than the optimal Kaiser-Bessel kernel,
while our modified ES kernel exceeds both other kernels in terms of accuracy.
\Cref{fig:comparekernels} again shows that it is possible to find a kernel shape with this code that leads to more accurate transforms than our modified ES kernel.
We also plot the spheroidal function kernel, which evidently performs much worse than the other kernels within the retained part of the image.
The comparison with this particular error function illustrates that the other kernels, which are chosen based on the knowledge that a part of the image will be trimmed, produce lower errors inside the final image in exchange for much higher errors in the trimmed regions.

Although the least-misfit kernel achieves a slightly more accurate gridding, we used the modified ES kernel for our implementation because only two real numbers are needed to specify the kernel for given $\alpha$ and $\sigma$ in contrast to much larger tables for the least-misfit kernel.
Additionally, it is non-trivial to minimise the supremum norm of \cref{eq:maperror} for a general least-misfit kernel.
With only two parameters, a brute force parameter search is affordable, but this does not work for the many more degrees of freedom of the least-misfit kernels.

\subsection{Kernel evaluation}\label{sec:kernel_eval}
In addition to choosing a kernel function that yields low errors, for the design of a practical algorithm it is also crucial to have a highly efficient way of evaluating this chosen function.
Because for every visibility processed it is necessary to evaluate the kernel at least 3$\alpha$ times ($\alpha$ times each in $u$-, $v$-, and $w$-direction), this is definitely a computational hot spot, and therefore a single evaluation should not take more than a few CPU cycles.

From the candidate functions listed in \cref{sec:kernelshape}, it is obvious that this rules out direct evaluation in most cases.
The only exception here is the original ES kernel (eq.~\ref{eq:semicircle}), which can be evaluated up to several hundred million times per second on a single CPU core using vector arithmetic instructions.
To retain high flexibility with respect to the choice of kernel function, some other approach is therefore needed.

Traditionally, this problem is often addressed using tables of precomputed function values evaluated at equidistant points, from which the desired kernel values are then obtained by interpolation.
Typically, zeroth-order (i.e.\ nearest-neighbour selection) and linear interpolation are used.

Interpolation at low polynomial degree soon leads to look-up tables that no longer fit into the CPU Level-1 and Level-2 caches when the required accuracy is increased, thus leading to high load on the memory subsystem, especially when running on multiple threads.
To overcome this, we adopted an approach presented by \citet{logsemicircle} and approximated the kernel in a piece-wise fashion by several higher order polynomials.
\citet{logsemicircle} reported that for a given desired accuracy $\epsilon$, it is sufficient to represent a kernel with support $\alpha$ by a set of $\alpha$ polynomials of degree $\alpha+3$.
This means that a kernel evaluation can be carried out using only $\alpha+3$ multiply-and-add instructions, and the total storage requirement for the polynomial coefficients is $\alpha(\alpha + 4)$ floating point numbers, which is negligible compared to the traditional look-up tables and much smaller than the CPU cache.

Because this approach is applicable to all kernel shapes discussed above, has sufficient accuracy (which can even be tuned by varying the degree of the polynomials), and has very low requirements on both CPU and memory, we used it in our implementation.
Details on the construction of the approximating polynomials are given in \cref{sec:kernel_impl}.

\section{Implementation}\label{sec:impl}

\subsection{Design goals and high-level overview}\label{sec:goals}
In order to make our code useful (and easy to use) in the broadest possible range of situations, we aim for the library to
have minimum external dependencies (to simplify installation),
have a minimum simple interface and be easily callable from different programming languages (to allow convenient use as a plug-in for existing radio-astronomical codes),
be well-suited for a broad range of problem sizes and required accuracies,
have a very low memory footprint for internal data structures,
and reach very high performance, but not at the cost of significant memory consumption.
We decided to provide the functionality as a component of the \texttt{ducc}\footnote{\url{https://gitlab.mpcdf.mpg.de/mtr/ducc}} collection of numerical algorithms.
Because this package already provides support for multi-threading, SIMD data types, FFTs, and all other algorithmic prerequisites, the code does not have external dependencies and only requires a compiler supporting the C++17 language standard.
Its interface only consists of two functions (to apply the gridding operator and its adjoint), which take a moderate number of parameters (scalars and arrays).
For illustration purposes, we list the interface documentation for the Python frontend of the library in \cref{app:interface}.
Similar to many other gridder implementations, the interface allows specifying individual weights for each visibility, as well as a mask for flagging arbitrary subsets of the measurement set; both of these parameters are optional, however (see \cref{app:interface}).
For an easy explicit understanding of the algorithm, we provide a compact Python and a slightly optimized Numpy and a Numba implementation of the $w$-gridding\footnote{\url{https://gitlab.mpcdf.mpg.de/mtr/ducc/-/blob/ducc0/python/demos/wgridder_python_implementations.py}}.

One important motivation for choosing C++ was its ability to separate the high-level algorithm structure from low-level potentially architecture-dependent implementation details.
As an example, while the algorithm is written only once, it is instantiated twice for use with single-precision and double-precision data types.
The single-precision version is faster, requires less memory, and may be sufficient for most applications, but the user may choose to use double precision in particularly demanding circumstances.
Similarly, advanced templating techniques allow us to make transparent use of vector arithmetic instructions available on the target CPU, be it SSE2, AVX, AVX2, FMA3/4, or AVX512F;\ this is invaluable to keep the code readable and easy to maintain.
The SIMD class of \texttt{ducc} supports the x86\_64 instruction set, but could be extended to other instruction sets (such as ARM64) if needed.

Especially due to the necessity of having a low memory footprint, the $w$-planes are processed strictly sequentially.
For the gridding direction (the degridding procedure is analogous), this means that for every $w$-plane,
all relevant visibilities are gridded onto the $uv$-grid, weighted accordingly to their $w$-coordinate,
the appropriate $w$-screen is applied to the grid,
the grid is transformed into the image domain via FFT,
and the resulting image is trimmed and added to the final output image.
This approach is arguably suboptimal from a performance point of view because it requires re-computing the kernel weights in $u$- and $v$-direction for every visibility at each $w$-plane it contributes to:
the number of kernel evaluations necessary to process a single visibility increases from $3\alpha$ to $\alpha(2\alpha+1)$.
On the other hand, processing several planes simultaneously would increase the memory consumption considerably, and at the same time the speed-up would probably not be very significant because kernel computation only accounts for a minor fraction of the overall run time ($\lessapprox 20$\%).

Overall, our approach requires the following auxiliary data objects:
a two-dimensional complex-valued array for the $uv$-grid (requiring $2\sigma^2$ times the size of the dirty image),
a temporary copy of the dirty image (only for degridding),
and a data structure describing the processing order of the visibilities (see \cref{sec:sorting} for a detailed description and a size estimate).
Processing only a single $w$-plane at a time implies that for parallelisation the relevant visibilities need to be subdivided into groups that are gridded or degridded concurrently onto/from that plane by multiple threads.
To obtain reasonable scaling with such an approach, it is crucial to process the visibilities in an order that is strongly coherent in $u$ and $v$;
in other words, visibilities falling into the same small patch in $uv$-space should be processed by the same thread and temporally close to each other.
This approach optimises both cache re-use on every individual thread as well as (in the gridding direction) minimising concurrent memory writes.
However, finding a close-to-optimal ordering for the visibilities in short time, as well as storing it efficiently, are nontrivial problems; they are discussed in \cref{sec:sorting}.

As mentioned initially, parameters for interferometric imaging tasks can vary extremely strongly:
the opening angle of the field of view can lie between arcseconds and $\pi$, visibility counts range from a few thousands to many billions, and image sizes start below $10^6$ pixels and reach $10^9$ pixels for current observations, with further increases in resolution to be expected.
Depending on the balance between these three quantities, the optimal choice (in terms of CPU time) for the kernel support $\alpha$, and depending on this the choice, of other kernel parameters and the oversampling factor $\sigma$, can vary considerably, and choosing these parameters badly can result in run times that are several times slower than necessary.
To avoid this, our implementation picks near-optimal $\alpha$ and $\sigma$ depending on a given task's parameters, based on an approximate cost model for the individual parts of the computation.
For all available $\alpha$ ($\alpha \in \{4, \ldots, 16\}$ in the current implementation), the code checks the list of available kernels for the one with the smallest $\sigma$ that provides sufficient accuracy and predicts the total run-time for this kernel using the cost model.
Then the kernel, $\alpha$, and $\sigma$ with the minimum predicted cost are chosen.

\subsection{Gridding kernel}\label{sec:kernel_impl}
Our code represents the kernel function by approximating polynomials as presented in \cref{sec:kernel_eval}.
A kernel with a support of $\alpha$ grid cells is subdivided into $\alpha$ equal-length parts, one for each cell, which are approximated individually by polynomials of degree $\alpha+3$.
When the kernel is computed in $u$- and $v$-directions, evaluation always takes place at locations spaced with a distance of exactly one grid cell, a perfect prerequisite for using vector arithmetic instructions.
As an example, for $\alpha=8$ and single precision, all eight necessary kernel values can be computed with only 11 FMA (floating-point multiply-and-add) machine instructions on any reasonably modern x86 CPU.

We used the family of modified ES kernels introduced in \cref{sec:kernelshape}.
They are convenient because an optimised kernel for given $\alpha$ and $\sigma$ is fully characterised by only two numbers $\beta$ and $\mu$, and therefore it is simple and compact to store a comprehensive list of kernels for a wide parameter range of $\alpha$, $\sigma$ and $\epsilon$ directly within the code.
This is important for the choice of near-optimal gridding parameters described in the preceding section.

When a kernel has been picked for a given task, it is converted to approximating polynomial coefficients.
For maximum accuracy, this should be done using the Remez algorithm \citep{remez},
but we found that evaluating the kernel at the respective Chebyshew points (for an expansion of degree $n$, these are the roots of the degree $(n+1)$ Chebyshev polynomial, mapped from $[-1; 1]$ to the interval in question) and using the interpolating polynomial through the resulting values produces sufficiently accurate results in practice while at the same time being much simpler to implement.
Chebyshew abscissas are used because the resulting interpolants are much less prone to spurious oscillations than those obtained from equidistant abscissas\footnote{\url{https://en.wikipedia.org/wiki/Runge\%27s_phenomenon}} \citep{runge-1901}.

Even better accuracy could be obtained by switching from modified ES kernels to least-misfit kernels, but there is a difficult obstacle to this approach:
determining a least-misfit kernel for a given $\alpha$ and $\sigma$, which is optimal in the supremum-norm sense instead of the L2-norm sense, may be possible only by a brute-force search, which may be unaffordably expensive.
Because the obtainable increase in accuracy is probably modest, we decided to postpone this improvement to a future improved release of the code.

\subsection{Optimising memory access patterns}\label{sec:sorting}
With the highly efficient kernel evaluation techniques described above, the pure computational aspect of gridding and degridding no longer dominates the run time of the algorithm.
Instead, most of the time is spent reading from and writing to the two-dimensional $uv$-grid.
Processing a single visibility requires $\alpha^3$ read accesses to this grid, and for the gridding direction, the same number of additional write accesses.
While it is not possible to reduce this absolute number without fundamentally changing the algorithm (which in turn will almost certainly lead to increasing complexity in other parts), much can be gained by processing the visibilities in a cache-friendly order, as was already pointed out in \cref{sec:goals}.
Making the best possible use of the cache is also crucial for good scaling behaviour because every CPU core has its own L1 and L2 caches, whereas there is only a small number of memory buses (with limited bandwidth) for the entire compute node. For multi-threaded gridding operations, this optimisation is even more important because it decreases the rate of conflicts between different threads trying to update the same grid locations;
without this measure, $R^\dagger$ would have extremely poor scaling behaviour.

Reordering the visibility and/or baseline data is not an option here because this would require either creating a rearranged copy of the visibilities (which consumes an unacceptable amount of memory) or, in the gridding direction, manipulating the input visibility array in-place (which is fairly poor interface design). Consequently, we rather used an indexing data structure describing the order in which the visibilities should be processed.

For this purpose, we subdivided the $uv$-grid into patches of $16\times 16$ pixels, which allowed us to assign a tuple of tile indices $(t_u, t_v)$ to every visibility.
The patch dimension was chosen such that for all supported $\alpha$ and arithmetic data types, the \enquote{hot} data set during gridding and degridding fit into a typical Level-1 data cache.
In $w$-direction, the index of the first plane onto which the visibility needs to be gridded is called $t_w$.
For compact storage, we used the fact that the $uvw$-locations of the individual frequency channels for a given row of the measurement set tend to be very close to each other.
In other words, it is highly likely that two visibilities that belong to the same row and neighbouring channels are mapped to the same $(t_u, t_v, t_w)$ tuple.

The resulting data structure is a vector containing all $(t_u, t_v, t_w)$ tuples that contain visibilities.
The vector is sorted lexicographically in order of ascending $t_u$, ascending $t_v$ , and finally ascending $t_w$.
Each of the vector entries contains another vector, whose entries are $(i_\text{row}, i_\text{chan,begin}, i_\text{chan,end})$ tuples, where $i_\text{row}$ is the row index of the visibility in question, and $i_\text{chan,begin}$ and $i_\text{chan,end}$ represent the first and one-after-last channel in the range, respectively. Each of these vectors is sorted lexicographically in order of ascending $i_\text{row}$ and ascending $i_\text{chan,begin}$.

While fairly nontrivial, this storage scheme is extremely compact: for a typical measurement set, it consumes roughly one bit per non-flagged visibility and is therefore much smaller than the visibility data themselves (which use eight bytes for every visibility, even the flagged ones).
In the most unfavourable case (which occurs, e.g., when the measurement set only contains a single channel or when every other frequency channel is flagged), the memory consumption will be around eight bytes per non-flagged visibility.

Processing the visibility data in this new ordering leads to a more random access pattern to the visibility array itself.
This is only a small problem, however, because entries for neighbouring channels are still accessed together in most cases, and also because the number of data accesses to the visibility array is lower by a factor of $\alpha^2$ than the one to the $uv$-grid in our algorithm.

\subsection{Parallelisation strategy}\label{sec:parallel}

Our code supports shared memory parallelisation by standard C++ threads, that is,\ it can be run on any set of CPUs belonging to the same compute node.
To achieve good scaling, all parts of the algorithm that contribute noticeably to the run time need to be parallelised.
In our case these parts are:
building the internal data structures,
performing the (de)gridding process,
applying $w$-screens,
evaluating Fourier transforms,
and evaluating and applying kernel corrections.

For the construction of the data structures (discussed in \cref{sec:sorting}), we subdivided the measurement set into small ranges of rows that are processed by the available threads in a first-come-first-serve fashion.
The threads concurrently update a global sorted data structure (using mutexes to guard against write conflicts), which is finally converted into the desired index list in a single-threaded code section.
While considerable speedups can be achieved by this approach compared to a purely single-threaded computation,
this part of the algorithm does not scale perfectly and can become a bottleneck at very high thread counts.

With the list of work items in hand, parallelising the actual gridding and degridding steps is straightforward:
first, the list is subdivided into a set of roughly equal-sized chunks with $n_{\text{chunks}}\gg n_{\text{threads}}$.
Each thread fetches the first chunk that has not been processed yet, performs the necessary operations, and then requests the next available chunk, until all chunks have been processed.
This kind of dynamic work balancing is required here because it is difficult to determine a priori how much CPU time a given chunk will require.

The way in which the list was constructed ensures that each chunk is confined to a compact region of the $uv$-plane and therefore reduces potential write conflicts between threads during gridding.
Still, it might happen that different threads try to update the same pixel in the $uv$-grid simultaneously, which would lead to undefined program behaviour.
To avoid this, each thread in both gridding and degridding routines employs a small buffer containing a copy of the $uv$-region it is currently working on, and when the gridding routine needs to write this back to the global $uv$-grid, this operation is protected with a locking mechanism.
In practice, the amount of time spent in this part of the code is very small, so that lock contention is not an issue.

Furthermore, the application of the $w$-screens and the kernel correction factors are parallelised by subdividing the array in question into equal-sized slabs, which are simultaneously worked on by the threads.
The FFT component has a built-in parallelisation scheme for multi-dimensional transforms that we make use of.

As mentioned above, the provided parallelisation can only be used on a single shared-memory compute node.
A further degree of parallelism can be added easily, for example by distributing the measurement set data evenly over several compute nodes, performing the desired gridding operation independently on the partial data sets, and finally summing all resulting images.
Analogously, for degridding, the image needs to be broadcast to all nodes first, and afterwards, each node performs degridding for its own part of the measurement set.
How exactly this is done will most likely depend on the particular usage scenario, therefore we consider distributed memory parallelisation to be beyond the scope of our library.
A distribution strategy over several compute nodes will increase the relative amount of time spent for computing the FFTs.
Still, our implementation partially compensates for this effect by picking a combination of $\alpha$, $\sigma$, and kernel shape that is optimised for the changed situation.

\section{Accuracy tests}\label{sec:validation}
This section reports the accuracy tests that we have performed to validate our implementation.
The tests can be subdivided into two major parts:
the accuracy with respect to the direct evaluation of the adjoint of \cref{eq:rime},
\begin{align}
(R_0^\dagger d)_{lm} = \frac{1}{n_{lm}}\, \sum_{k} e^{2\pi \iu [u_k l + v_k m - w_k(n_{lm}-1)]} \, d_k, \quad l\in L, m\in M,
\end{align}
and the adjointness consistency between the forward and backward direction of the different calls.

\subsection{Adjointness consistency}\label{sec:adjointness}
First, the degridding and the gridding calls were checked for their consistency.
This is possible because mathematically, the two calls are the adjoint of each other.
Therefore
\begin{align}
 \Re\bigg(\Big\langle R(I),d \Big\rangle_{(1)}\bigg) \overset{!}{=}\Big \langle I, R^\dagger (d)  \Big\rangle_{(2)} , \quad \forall\, I\in\mathbb R^{n_ln_m}, \forall\, d \in \mathbb C^{n_k},
  \label{eq:adjointness}
\end{align}
where $\langle a, b\rangle_{(1)} \coloneqq a^\dagger b$ and $\langle a, b\rangle_{(2)} \coloneqq a^T b$ are the dot products of $\mathbb C^{n_k}$ and $\mathbb R^{n_l n_m}$, respectively.
On the left-hand side of the equation, the real part needs to be taken because $R$ maps from an $\mathbb R$- to a $\mathbb C$-vector space.
Still, $\Im (R(I))$ is tested by \cref{eq:adjointness} because evaluating the scalar product involves complex multiplications.
Therefore the real part of the scalar product also depends on $\Im (R(I))$.

For the numerical test, we chose $n_l= n_m = 512$ and a field of view of $15^\circ \times 15^\circ$.
The observation was performed at 1~GHz with one channel.
The synthetic $uvw$-coverage consisted of 1000~points sampled from a uniform distribution in the interval $[-a, a]$, where $a=\mathrm{pixsize}/2/\lambda$, pixsize is the length of one pixel and $\lambda$ is the observing wave length.
The real and the imaginary parts of the synthetic visibilities $d$ were drawn from a uniform distribution in the interval $[-0.5,0.5]$.
Analogously, we drew the pixel values for the dirty image $I$ from the same distribution.
We consider this setup to be generic enough for accuracy testing purposes.

As discussed above, our implementation supports applying or ignoring the $w$-correction and can run in single or double precision.
This gives four modes that are tested individually in the following.
Moreover, the kernel sizes and the oversampling factor were chosen based on the intended accuracy $\epsilon$, specified by the user.
As a criterion for the quality of the adjointness, we use
\begin{equation}
\epsilon_\text{adj} \coloneqq \frac{\left| \Re\left(\Big\langle R(I),d \Big\rangle_{(1)}\right) - \left \langle I, R^\dagger (d)  \right\rangle_{(2)}\right|}
                            {\min{\left(\norm{d}\cdot\norm{R(I)},
                            \norm{I} \cdot \norm{R^\dagger(d)}\right)}}\text{.}
\end{equation}
For all four modes and for all tested $\epsilon$ in the supported range ($\geq 10^{-5}$ for single precision, $\geq 10^{-14}$ for double precision), this quantity lay below $10^{-7}$ and $10^{-15}$ for single and double precision, respectively.

\subsection{Accuracy of $R^\dagger$}
Second, we compared the output of our implementation to the output of the direct Fourier transform with and without $w$-correction.
It suffices to test only $R^\dagger$ and not also $R$ because the consistency of the two directions was already verified in \cref{sec:adjointness}.
The error is quantified as rms error,
\begin{align}
  \epsilon_{\mathrm{rms}}(d) = \sqrt{\frac{\sum_{lm} \left| (R_0^\dagger d)_{lm}-(R^\dagger d)_{lm} \right|^2}{\sum_{lm} \left| (R_0^\dagger d)_{lm} \right|^2}}.
\end{align}
\begin{figure}
  \centering
  \input{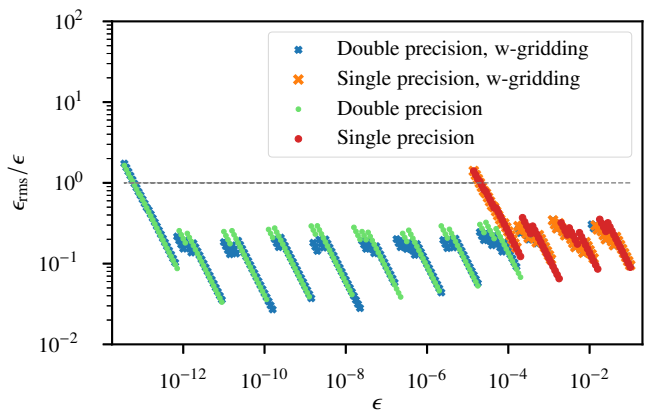}
  \caption{Accuracy of $R^\dagger$.
    The ratio of measured root mean square error to the requested accuracy $\epsilon$ is plotted as a function of $\epsilon$ itself.
    The grey line denotes the identity function.
    Points lying in the region below the line represent configurations that are more accurate than specified by the user.
  }\label{fig:accuracy}
\end{figure}
As testing setup, the same configuration as above was employed.
\Cref{fig:accuracy} shows the results of the (approximate) gridding implementation against the exact DFT.
It is apparent that single precision transforms reach the requested accuracy for $\epsilon\gtrapprox 3\cdot 10^{-5}$, while double precision transforms are reliably accurate down to $\epsilon\approx10^{-13}$.
We deliberately also show results for $\epsilon$ outside this safe range to demonstrate how the resulting errors grow beyond the specified limit due to the intrinsic inaccuracy of floating-point arithmetics.
Inside the safe region, the achieved accuracy typically lies in the range between $0.03\epsilon$ and $\epsilon$, which indicates that the estimation in \cref{eq:cauchyschwarz} is not overly pessimistic.

The saw-tooth pattern of the measured errors is caused by the dynamic parameter selection during the setup process of each gridding operation mentioned near the end of \cref{sec:goals}:
Moving from higher to lower accuracies, a fixed combination of $\alpha$, $\sigma$, and the corresponding kernel shape results in decreasing $\epsilon_\text{rms}/\epsilon$, which is indicated by the individual descending line segments.
At some point, a new parameter combination (lower $\sigma$, or lower $\alpha$ with increased $\sigma$) with sufficiently high accuracy and lower predicted run time becomes available.
This is then selected and the relative error jumps upwards, while still remaining well below the specified tolerance.

\section{Performance tests}\label{sec:benchmarks}
The tests in this section were performed on a 12-core AMD Ryzen 9 3900X CPU with 64GB main memory attached.
\texttt{g++} 10.2 was used to compile the code, with notable optimisation flags including \texttt{-march=native}, \texttt{-ffast-math}, and \texttt{-O3}.
The system supports two hyper-threads per physical CPU core, so that some of the tests were executed on up to 24 threads.
As test data we used a MeerKAT \citep{meerkat} L-band measurement set corresponding to an 11-hour synthesis with \SI{8}{\second} integration time and 2048 frequency channels, using 61 antennas (824476 rows in total, project id 20180426-0018).
We worked on the sum of XX and YY correlations only, ignoring polarisation, and after selecting only unflagged visibilities with non-vanishing weights, roughly 470 million visibilities need to be processed for each application of the gridding or degridding operator.
The size of the dirty image was $4096\times 4096$ pixels, and the specified field of view was $1.6^\circ\times1.6^\circ$.
Unless mentioned otherwise, computations were executed in single-precision arithmetic and with a requested accuracy of $\epsilon=10^{-4}$.
We compared the timings of our implementation to the standard radio software \texttt{wsclean} and the general-purpose library \texttt{FINUFFT}\footnote{\url{https://github.com/flatironinstitute/finufft}, type 1, two-dimensional transform.}.

\subsection{Strong scaling}\label{sec:strong_scaling}

\begin{figure}
  \centering
  \input{img/strong_scaling.pgf}
  \caption{Strong-scaling scenario.
    The vertical dotted gray line indicates the number of physical cores on the benchmark machine.
    Efficiency is the theoretical wall time with perfect scaling divided by the measured wall time and divided by the single-thread timing of \enquote{$R^\dagger$ ducc}.
  }
  \label{fig:strongscaling}
\end{figure}

First, we investigated the strong-scaling behaviour of our implementation.
\Cref{fig:strongscaling} shows the timings of this problem evaluated with a varying number of threads.
The ideal scaling would of course be $\propto n_\text{threads}^{-1}$, but this cannot be fully reached in practice.
As mentioned in \cref{sec:parallel}, the setup part of the algorithm does not scale perfectly, and the same is true for the FFT operations because of their complicated and not always cache-friendly memory access patterns.

Still, the implementation scales acceptably well, reaching a speed-up of roughly \input{numbers_speedup_gridding0} when running on 12~threads.
While the further improvements are much lower when scaling beyond the number of physical cores, as has to be expected, a total speed-up of around \input{numbers_speedup_gridding1} is reached when using all hyper-threads available on the system.

In this test, degridding is slightly, but consistently slower than gridding, which appears counter-intuitive because degridding only requires roughly half the number of memory accesses.
We assume that this is due to the horizontal addition of vector registers that has to be performed when a computed visibility value is written back to the measurement set.
This kind of operation is notoriously slow on most CPUs, while the corresponding broadcast operation that is needed during gridding is much faster.
If this interpretation is correct, it indicates that in the selected regime (single precision with an accuracy of $10^{-4}$) memory accesses do not completely dominate computation.
For higher accuracies this is no longer true, as shown in \cref{sec:epsilon_scaling}.

\Cref{fig:strongscaling} also shows analogous timings for the standard gridder in \texttt{wsclean}, but it is important to note that these cannot be directly compared to those of our code.
While we tried to measure the timings with as little overhead as possible (we used the times reported by \texttt{wsclean} itself for the operations in question), the \texttt{wsclean} default gridder always interleaves I/O operations (which do not contribute at all to our own measurements) with the actual gridding and degridding, so there is always an unknown, non-scaling amount of overhead in these numbers.
Additionally, the accuracy of \texttt{wsclean} cannot be set explicitly; based on experience, we expect it to be close to the target of $10^{-4}$ near the image center, but somewhat worse in the outer regions.

\begin{figure}
  \centering
  \input{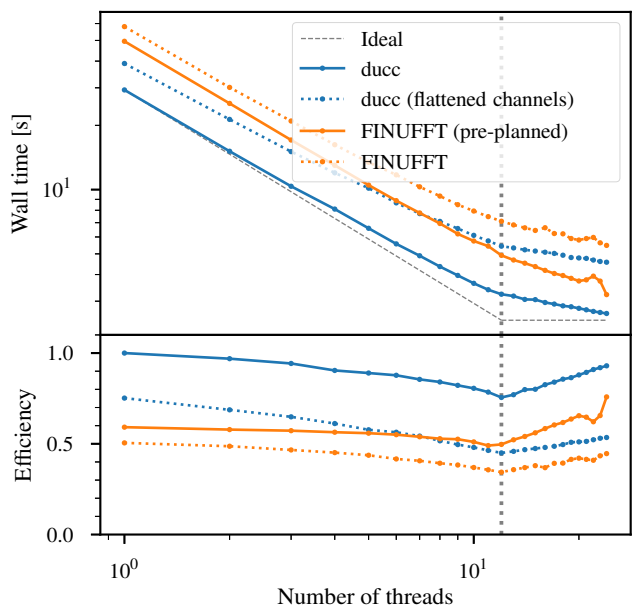}
  \caption{Comparison to FINUFFT.
    The vertical dotted grey line indicates the number of physical cores on the benchmark machine.
    Efficiency is the theoretical wall time with perfect scaling divided by the measured wall time and divided by the single-thread timing of \enquote{ducc}.
  }
  \label{fig:finufft}
\end{figure}

\subsection{Comparison to non-equidistant FFT}
As mentioned in the introduction, gridding or degridding without the $w$-term can be interpreted as a special case of the non-uniform FFT, where the $uv$ coordinates of the individual points are not independent, but vary linearly with frequency in each channel.
For this reason we also performed a direct comparison of our implementation with the \texttt{FINUFFT} library \citep{logsemicircle}.
We still used the same measurement set as above, but performed a gridding step without the $w$ term, using double precision and requiring $\epsilon=10^{-10}$.

Because a general non-uniform FFT algorithm cannot be informed about the special structure of the $uv$ coordinates, we supplied it with an explicit coordinate pair for every visibility.
This implies that a much larger amount of data is passed to the implementation, and it also increases the cost of the preprocessing step.
To allow a fairer comparison, we also ran \texttt{ducc} on an equivalent flattened data set, which only contained a single frequency channel and therefore as many $uv$ coordinates as there are visibilities.
We verified that both implementations returned results that are equal to within the requested tolerance.
The performance results are shown in \cref{fig:finufft}. In contrast to our implementation, \texttt{FINUFFT} features a separate planning phase that can be timed independently, so we show \texttt{FINUFFT} timings with and without the planning time, in addition to \texttt{ducc} timings for processing the original and flattened measurement set.

To a large extent, the results confirm the expectations.
\texttt{FINUFFT} is always slower than \texttt{ducc} when \texttt{ducc} works on the un-flattened data.
This can be attributed to the slightly higher accuracy of the \texttt{ducc}  kernels and/or to its advantage of knowing the internal structure of the $uv$ data, which reduces setup time and the amount of memory accesses considerably.
Furthermore, \texttt{ducc} performs rather poorly on the flattened data compared to its standard operation mode, especially with many threads.
Here it becomes obvious that the index data structure, which has many benefits for multi-channel data, slows the code down when it is not used as intended by providing only a single channel.
Finally, pre-planned \texttt{FINUFFT} performs worse than \texttt{ducc} with flattened data at low thread counts, but has a clear speed advantage on many threads;
again, this is probably due to the \texttt{ducc} data structures, which are suboptimal for this scenario.

Memory consumption also behaves as expected, meaning that \texttt{ducc} without flattening requires the least amount of memory (because it does not need to store the redundant $uv$ data), followed by both \texttt{FINUFFT} runs, while \texttt{ducc} with flattening consumes the most memory because it stores the full $uv$ coordinates as well as a really large index data structure.
Overall, we consider it very encouraging that despite differences in details, the performance and scaling behaviour of these two independent implementations are fairly similar to each other.

\subsection{Run time vs.\ accuracy}\label{sec:epsilon_scaling}
\begin{figure}
  \centering
  \input{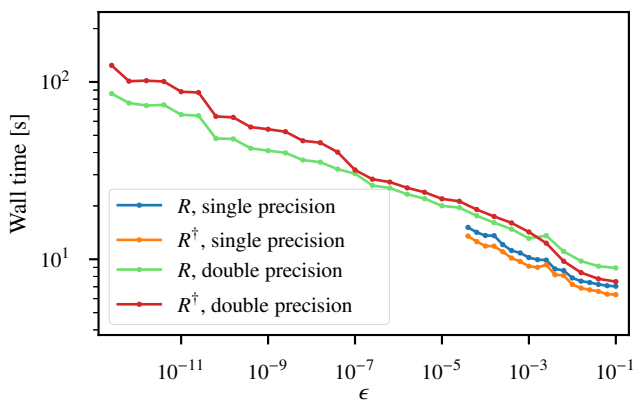}
  \caption{Wall time vs.\ specified accuracy $\epsilon$ measured with six threads.}
  \label{fig:epsilon}
\end{figure}

For the following tests, we again used the setup described at the beginning of this section, but we fixed the number of threads to six and varied the requested accuracy $\epsilon$ as well as the data type of the input (single or double precision).
\Cref{fig:epsilon} shows the expected decrease in wall time for increasing $\epsilon$, that is, lower accuracy.
In single-precision mode the evaluation is indeed slightly faster than in double precision, most probably because more visibilities and grid points can be communicated per second between CPU and RAM for a given memory bandwidth.
Moreover, the number of elements in the CPU vector registers is twice as large for single-precision variables.

In analogy to the observations in \cref{sec:strong_scaling}, degridding is slightly slower than gridding for these measurements.
For double precision, the same is only true at very low accuracies; for $\epsilon\lessapprox 10^{-3}$, gridding becomes the more expensive operation, and this trend becomes very pronounced at the lowest reachable $\epsilon$ values.
In these runs, the kernel support $\alpha$ is quite large and most of the run-time is presumably spent on data transfer from/to main memory.
The results also show that while certainly attainable, high accuracy comes at a significant cost:
going from a typical $\epsilon$ of $10^{-4}$ to $10^{-12}$ increases the run-time by about an order of magnitude.

\balance
\section{Discussion}\label{sec:conclusion}
We have presented a new implementation of the radio interferometry gridding and degridding operators, which combines algorithmic improvements from different sources:
an accurate and efficient treatment of the $w$-term for wide-field observations published by \citet{wgridding,ye21},
an easy-to-use, high-accuracy, functional form for the gridding kernels presented by \citet{logsemicircle}, with some slight improvements,
a piecewise polynomial approximation method for arbitrary kernels (also published by \citealt{logsemicircle}), which is very well suited for the task at hand),
a parallelisation strategy, dynamic parameter selection, and indexing data structure of our own design.
To the best of our knowledge, the resulting code compares favourably to other existing Fourier-domain gridders (both for wide- and narrow-field data) in terms of accuracy, memory consumption, single-core performance, and scalability.
Our implementation is designed to have minimum dependencies (only a C++17 compiler is needed), and it is free and open-source software.
Therefore it may be advantageous to add it as an alternative option to existing radio interferometry imaging packages, as was already done in the \texttt{wsclean} code.

Compared with the fairly recent image-domain gridding approach \citep[IDG,][]{idg}, it appears that our implementation currently has a performance advantage when both algorithms are run on CPUs, but the GPU implementation of IDG easily outperforms all competitors on hardware of comparable cost.
Furthermore, IDG can incorporate direction-dependent effects (DDEs) in a straightforward manner, which are difficult and costly to treat with Fourier-domain gridding algorithms.

However, it may be possible to address this within the $w$-gridding framework.
The A-stacking algorithm \citep{astacking} might be combined with $w$-gridding, for instance.
This would imply
approximating all possible DDE patterns as linear combinations of a small set of $N_b$ basis functions $f_b(l,m)$,
computing (for every visibility) the projection of its particular DDE pattern onto this set of functions,
running the $w$-gridder $N_b$ times with the appropriate sets of weights,
multiplying each result with the corresponding basis function, and finally adding everything together.
Investigating the actual feasibility and performance of such an approach is left for future studies.

\begin{acknowledgements}
  We thank Landman Bester, Simon Perkins, Wasim Raja and Oleg Smirnov for testing and feedback on the interface,
  Alex Barnett, Vincent Eberle, Torrance Hodgson and Haoyang Ye for feedback on drafts of the manuscript,
  and SARAO for providing access to MeerKAT data for our algorithmic testing purposes.
  Philipp Arras acknowledges financial support by the German Federal Ministry of Education and Research (BMBF) under grant 05A17PB1 (Verbundprojekt D-MeerKAT).
\end{acknowledgements}

\bibliographystyle{aa}
\bibliography{bib.bib}


\appendix
\nobalance

\section{Kernel parameters}\label{app:kernelparams}

Optimal kernel parameters and associated accuracy $\epsilon$ for the modified exponential semicircle kernel (eq.\ \ref{eq:semicircle2}) given the oversampling factor $\sigma$ and the kernel support size $\alpha$.
Larger $\sigma$ and larger $\alpha$ lead to smaller $\epsilon$.
Larger $\sigma$ and smaller $\alpha$ increase the fraction of the FFT of the total computation time.
FFT and gridding costs are represented in our implementation with a simple cost model such that the algorithm can choose optimal $\alpha$ and $\sigma$ automatically.
For brevity, we display only the tables for $\alpha \in \{4, 7, 8, 12, 16\}$.
The rest can be looked up in the \texttt{ducc} code repository.
The least-misfit kernels \citep{optimalgridding} achieve an accuracy $\epsilon = 10^{-7}$ for $\alpha = 7$ and $\sigma = 2$.

\input{table_modifiedeskernel}

\clearpage

\section{Python interface documentation}\label{app:interface}
{\footnotesize
\begin{verbatim}
def ms2dirty(uvw, freq, ms, wgr, npix_x, npix_y,
             pixsize_x, pixsize_y, nu, nv, epsilon,
             do_wstacking, nthreads, verbosity, mask):
"""
Converts an MS object to dirty image.

Parameters
----------
uvw: numpy.ndarray((nrows, 3), dtype=numpy.float64)
    UVW coordinates from the measurement set
freq: numpy.ndarray((nchan,), dtype=numpy.float64)
    channel frequencies
ms: numpy.ndarray((nrows, nchan),
    dtype=numpy.complex64 or numpy.complex128)
    the input measurement set data.
    Its data type determines the precision in which
    the calculation is carried out.
wgt: numpy.ndarray((nrows, nchan), float with same
    precision as `ms`), optional
    If present, its values are multiplied to the
    input before gridding.
npix_x, npix_y: int
    dimensions of the dirty image (must both be even
    and at least 32)
pixsize_x, pixsize_y: float
    angular pixel size (in radians) of the dirty image
nu, nv: int
    obsolete, ignored
epsilon: float
    accuracy at which the computation should be done.
    Must be larger than 2e-13. If `ms` has type
    numpy.complex64, it must be larger than 1e-5.
do_wstacking: bool
    if True, the full w-gridding algorithm is carried
    out, otherwise the w values are assumed to be zero
nthreads: int
    number of threads to use for the calculation
verbosity: int
    0: no output
    1: some output
    2: detailed output
mask: numpy.ndarray((nrows, nchan),
                    dtype=numpy.uint8),
    optional
    If present, only visibilities are processed
    for which mask!=0

Returns
-------
numpy.ndarray((npix_x, npix_y), dtype=float of same
    precision as `ms`)
    the dirty image

Notes
-----
The input arrays should be contiguous and in C memory
order. Other strides will work, but can degrade
performance significantly.
"""
\end{verbatim}
\newpage

\begin{verbatim}

def dirty2ms(uvw, freq, dirty, wgr, pixsize_x,
             pixsize_y, nu, nv, epsilon, do_wstacking,
             nthreads, verbosity, mask):
"""
Converts a dirty image to an MS object.

Parameters
----------
uvw: numpy.ndarray((nrows, 3), dtype=numpy.float64)
    UVW coordinates from the measurement set
freq: numpy.ndarray((nchan,), dtype=numpy.float64)
    channel frequencies
dirty: numpy.ndarray((npix_x, npix_y),
    dtype=numpy.float32 or numpy.float64)
    dirty image
    Its data type determines the precision in which
    the calculation is carried out.
    Both dimensions must be even and at least 32.
wgt: numpy.ndarray((nrows, nchan), same dtype as
   `dirty`), optional
    If present, its values are multiplied to the
    output.
pixsize_x, pixsize_y: float
    angular pixel size (in radians) of the dirty image
nu, nv: int
    obsolete, ignored
epsilon: float
    accuracy at which the computation should be done.
    Must be larger than 2e-13.
    If `dirty` has type numpy.float32, it must be
    larger than 1e-5.
do_wstacking: bool
    if True, the full w-gridding algorithm is carried
    out, otherwise the w values are assumed to be zero
nthreads: int
    number of threads to use for the calculation
verbosity: int
    0: no output
    1: some output
    2: detailed output
mask: numpy.ndarray((nrows, nchan),
                    dtype=numpy.uint8),
    optional
    If present, only visibilities are processed
    for which mask!=0

Returns
-------
numpy.ndarray((nrows, nchan), dtype=complex of same
    precision as `dirty`)
    the measurement set data.

Notes
-----
The input arrays should be contiguous and in C memory
order. Other strides will work, but can degrade
performance significantly.
"""
\end{verbatim}
}
\end{document}

%% file: maperrorsupport.tex
6

%% file: mesvskb_alpha.tex
6

%% file: mesvskb_sigma.tex
1.5

%% file: numbers_speedup_gridding0.tex
8.0%

%% file: numbers_speedup_gridding1.tex
9.6%

%% file: table_modifiedeskernel.tex
\begin{table}[H] \centering\begin{tabular}{llll}
\hline\hline $\sigma$ & $\epsilon$ & $\beta$ & $\mu$ \\\hline
1.15 & 0.025654879 & 1.3873426689 & 0.5436851297\\
1.2 & 0.013809249 & 1.3008419165 & 0.5902137484\\
1.25 & 0.0085840685 & 1.3274088935 & 0.5953499486\\
1.3 & 0.0057322498 & 1.3617063353 & 0.5965631622\\
1.35 & 0.0042494419 & 1.384549988 & 0.5990241291\\
1.4 & 0.0033459552 & 1.4405325088 & 0.5924776015\\
1.45 & 0.0028187359 & 1.4635220066 & 0.5929442711\\
1.5 & 0.0023843943 & 1.5539689162 & 0.5772217314\\
1.55 & 0.0020343796 & 1.5991008653 & 0.5721765215\\
1.6 & 0.0017143851 & 1.6581546365 & 0.5644747137\\
1.65 & 0.0014730848 & 1.7135331415 & 0.5572788589\\
1.7 & 0.0012554492 & 1.7464330378 & 0.5548742415\\
1.75 & 0.0010610904 & 1.7887326906 & 0.5509877716\\
1.8 & 0.00090885567 & 1.8122309426 & 0.5502273972\\
1.85 & 0.0007757401 & 1.8304451327 & 0.550396716\\
1.9 & 0.0006740398 & 1.8484487383 & 0.5502376937\\
1.95 & 0.00058655391 & 1.8742215688 & 0.5489738941\\
2.0 & 0.00051911189 & 1.90694363 & 0.5468009434\\
\hline
\end{tabular}\caption{Optimal parameters for $\alpha = 4$.}\end{table}\begin{table}[H] \centering\begin{tabular}{llll}
\hline\hline $\sigma$ & $\epsilon$ & $\beta$ & $\mu$ \\\hline
1.15 & 0.00078476028 & 1.5248706519 & 0.5288306317\\
1.2 & 0.00027127166 & 1.5739348793 & 0.5287992619\\
1.25 & 0.00012594628 & 1.6245240723 & 0.527921777\\
1.3 & 7.0214545e-05 & 1.6835745981 & 0.5257484101\\
1.35 & 4.1972457e-05 & 1.7343424414 & 0.5239793844\\
1.4 & 2.378019e-05 & 1.7845017738 & 0.5224266045\\
1.45 & 1.3863408e-05 & 1.8180597789 & 0.5221834768\\
1.5 & 9.1605353e-06 & 1.868082272 & 0.5206277502\\
1.55 & 6.479159e-06 & 1.9188980015 & 0.5183134674\\
1.6 & 4.6544571e-06 & 1.9536166143 & 0.5178695891\\
1.65 & 3.5489761e-06 & 1.9786267068 & 0.5178430252\\
1.7 & 2.7030348e-06 & 2.0027666534 & 0.5178577604\\
1.75 & 2.0533894e-06 & 2.0289949199 & 0.5176300336\\
1.8 & 1.6069122e-06 & 2.0596412946 & 0.5167551932\\
1.85 & 1.2936794e-06 & 2.0720606842 & 0.5178747891\\
1.9 & 1.0768664e-06 & 2.090898174 & 0.5181009847\\
1.95 & 9.0890421e-07 & 2.1086185697 & 0.5184537843\\
2.0 & 7.7488775e-07 & 2.1278284187 & 0.5186377792\\
\hline
\end{tabular}\caption{Optimal parameters for $\alpha = 7$.}\end{table}\begin{table}[H] \centering\begin{tabular}{llll}
\hline\hline $\sigma$ & $\epsilon$ & $\beta$ & $\mu$ \\\hline
1.15 & 0.00026818611 & 1.568124649 & 0.5223052481\\
1.2 & 7.8028732e-05 & 1.620926145 & 0.5219287175\\
1.25 & 2.7460918e-05 & 1.6851585171 & 0.519925059\\
1.3 & 1.3421658e-05 & 1.7442373315 & 0.5182155619\\
1.35 & 7.5158217e-06 & 1.7876782642 & 0.5176319503\\
1.4 & 4.2472384e-06 & 1.8294321912 & 0.5171860211\\
1.45 & 2.5794802e-06 & 1.871691821 & 0.5161733611\\
1.5 & 1.6131994e-06 & 1.9213040541 & 0.5145350888\\
1.55 & 1.0974814e-06 & 1.9637229131 & 0.5134005827\\
1.6 & 7.531955e-07 & 2.0002761373 & 0.5128849282\\
1.65 & 5.5097346e-07 & 2.0275645736 & 0.5127082324\\
1.7 & 4.0136726e-07 & 2.0498410409 & 0.5130237662\\
1.75 & 2.906467e-07 & 2.073158517 & 0.5131757153\\
1.8 & 2.1834922e-07 & 2.0907418726 & 0.5136046561\\
1.85 & 1.6329905e-07 & 2.1164552354 & 0.5133333878\\
1.9 & 1.2828598e-07 & 2.126157016 & 0.5143004427\\
1.95 & 1.0171134e-07 & 2.1363206613 & 0.515235491\\
2.0 & 8.1881369e-08 & 2.1397013368 & 0.5166895497\\
\hline
\end{tabular}\caption{Optimal parameters for $\alpha = 8$.}\end{table}\begin{table}[H] \centering\begin{tabular}{llll}
\hline\hline $\sigma$ & $\epsilon$ & $\beta$ & $\mu$ \\\hline
1.15 & 2.7535895e-06 & 1.6661837519 & 0.5098172147\\
1.2 & 5.2570038e-07 & 1.7294557459 & 0.5089239596\\
1.25 & 1.378658e-07 & 1.7698182384 & 0.5099240718\\
1.3 & 4.4329167e-08 & 1.8092042442 & 0.510607427\\
1.35 & 1.7038991e-08 & 1.8619112597 & 0.5093832337\\
1.4 & 6.5438748e-09 & 1.9069147481 & 0.5089479889\\
1.45 & 2.9874764e-09 & 1.9318398074 & 0.5098082325\\
1.5 & 1.4920459e-09 & 1.9628483155 & 0.5100985753\\
1.55 & 8.0989276e-10 & 2.0129847811 & 0.5085327805\\
1.6 & 4.1660575e-10 & 2.0517921747 & 0.5079102398\\
1.65 & 2.3539727e-10 & 2.06983884 & 0.5085131064\\
1.7 & 1.3497289e-10 & 2.0887365361 & 0.5090417146\\
1.75 & 8.3256938e-11 & 2.106955733 & 0.5095920671\\
1.8 & 5.8834619e-11 & 2.1359415217 & 0.5091887069\\
1.9 & 2.6412908e-11 & 2.2006369514 & 0.5075889699\\
1.95 & 1.7189689e-11 & 2.2146741638 & 0.5080017404\\
2.0 & 1.2174796e-11 & 2.2431392199 & 0.5075191177\\
\hline
\end{tabular}\caption{Optimal parameters for $\alpha = 12$.}\end{table}\begin{table}[H] \centering\begin{tabular}{llll}
\hline\hline $\sigma$ & $\epsilon$ & $\beta$ & $\mu$ \\\hline
1.3 & 1.1509596e-10 & 1.7892839755 & 0.5122877693\\
1.35 & 3.2440049e-11 & 1.8914441282 & 0.5063521839\\
1.4 & 8.4329616e-12 & 1.9296369098 & 0.5065170208\\
1.45 & 3.1161739e-12 & 1.9674735425 & 0.5063244338\\
1.5 & 1.2100308e-12 & 2.0130787701 & 0.5055587965\\
1.55 & 4.6082202e-13 & 2.0438032614 & 0.5056309683\\
1.6 & 1.7883238e-13 & 2.0329561822 & 0.5089045671\\
1.65 & 9.2853815e-14 & 2.0494514743 & 0.5103582604\\
1.7 & 5.6614567e-14 & 2.0925119791 & 0.5083767402\\
1.75 & 2.875391e-14 & 2.1461524027 & 0.5062037834\\
1.8 & 1.6578982e-14 & 2.1490040175 & 0.508272183\\
1.85 & 1.1782751e-14 & 2.1811826814 & 0.5072570059\\
1.9 & 8.9196865e-15 & 2.1981176583 & 0.5075840871\\
1.95 & 6.6530006e-15 & 2.234001135 & 0.5060133105\\
2.0 & 5.0563492e-15 & 2.2621631913 & 0.5056924675\\
\hline
\end{tabular}\caption{Optimal parameters for $\alpha = 16$.}\end{table}